\documentclass[aps,prc,preprint,groupedaddress,preprintnumbers,
amsmath,amssymb]{revtex4}
\usepackage{graphicx}
\usepackage{bm}% bold math

\begin{document}

\title{Ab initio calculation of the $^4$He$(e,e'd)d$ reaction}

\author{Diego Andreasi$^{1,2}$, Sofia Quaglioni$^{1,2}$\footnote{present address:
Department of Physics, P.O. Box 210081, University of Arizona, Tucson, Arizona
85721}, Victor D. Efros$^3$, 
Winfried Leidemann$^{1,2}$, and Giuseppina Orlandini$^{1,2}$}

\affiliation{$^1$Dipartimento di Fisica, Universit\`a di Trento, I-38050 Povo, Italy}
 
\affiliation{$^2$Istituto Nazionale di Fisica Nucleare, Gruppo Collegato di Trento, 
I-38050 Povo, Italy}

\affiliation{$^3$Russian Research Centre "Kurchatov Institute",  123182 Moscow,  Russia}

\date{\today}

\begin{abstract}
The two--body knock--out reaction $^4$He$(e,e'd)d$ is calculated at various momentum 
transfers. The full four--nucleon dynamics is taken into account 
microscopically both in the initial and the final states. 
%The calculation of the final state interaction (FSI) is carried
%out with the Lorentz Integral Transform method. 
As NN interaction the central MT-I/III potential is used. The calculation shows 
a strong reduction of the coincidence cross section due to the final state interaction.
Nonetheless the theoretical results exhibit a considerable overestimation of 
the experimental cross section at lower momentum transfer. Comparisons with other, 
less complete, calculations suggest that  consideration of a more realistic 
ground state might not be sufficient for a good agreement with 
experiment, rather a more realistic final state interaction could play an essential 
role.  
\end{abstract}

\maketitle

\section{INTRODUCTION}
\label{Sec:Introduction}

Electron induced two--body knock--out reactions are considered to be an 
important tool to investigate nucleon--nucleon (NN) correlations in nuclei.
In comparison to one--nucleon knock--out processes two--nucleon emission reactions 
give more detailed information on NN dynamics, while many important details 
are already integrated out in the former case.  Hence quite a number of 
experimental and theoretical studies have been devoted to $(e,e'$NN) reactions 
in order to study NN  correlations (see e.g. \cite{overview}). However, it is often not very easy 
to obtain a clear picture of the two--body correlations, in fact one would need 
theoretical calculations where all relevant effects contributing to the observable 
under investigation are taken into account. For this reason in many theoretical 
works also effects from final state interaction (FSI), as well as meson exchange and 
$\Delta$ currents have been considered. Microscopic calculations 
of two--body knock--out reactions with the proper final continuum state have only 
been carried out for two-- and three--nucleon systems (see e.g. \cite{ALT,BoKr}). 
In more complex nuclei 
such exact and consistent studies are still missing.

The aim of the present paper is as follows. We want to consider 
the full FSI in a two--body emission reaction microscopically and at the same time
go beyond the above mentioned two-- and three--nucleon systems. To this
end we consider the $^4$He$(e,e'd)d$ reaction. 
The choice of this particular reaction is based on three different considerations.
The first is that, differently from the two-- and three--body systems, $^4$He has 
some characteristics (binding energy per nucleon, density) rather similar to those 
of heavier nuclei.
Secondly, the $(e,e'd)$ reaction has been suggested as a particularly useful
tool for investigating short range correlations (for a
brief summary see \cite {LOTT2}). In fact the $^4$He$(e,e'd)d$ reaction was 
among the first NIKHEF experiments dedicated to the study of NN correlations 
\cite{NIKHEF}. The third reason is that in ref. \cite{LOTT1} large 
effects of FSI were found in a two--deuteron cluster model. 

In the present paper
we perform a rigorous calculation of the FSI of the four--nucleon system, where the
nucleons interact via an NN potential. We 
employ an integral transform method as outlined in ref. \cite{Efros}. Particularly 
suited for such kind of calculations is the Lorentz Integral Transform (LIT), which 
was proposed in ref. \cite{LIT}. In fact the LIT has already been used 
for the calculation
of various exclusive reactions: $d(e,e'p)n$ \cite{Lapiana}, $^4$He$(\gamma,p)^3$H 
\cite{Sofia1}, $^4$He$(\gamma,n)^3$He \cite{Sofia1} and $^4$He$(e,e'p)^3$H
\cite{Sofia2}.

Our paper is organized as follows.
In sect.~2 we describe the formalism  of the $^4$He$(e,e'd)d$
cross section. A discussion of results and a short conclusion are given in sect.~3.

\section{formalism}

The longitudinal part of the $^4$He$(e,e'd)d$ coincidence cross section 
is given by
\begin{equation}         
\label{cross}                                            
\frac{d^5\sigma_L}{de_{k^\prime} d\Omega_{e^\prime} d\Omega_d} 
=\sigma_M \, \frac{M_d p_d}{2 (1-(q/2p_d) \cos(\theta_d))} 
 \, \frac{q_\mu^4}{q^4}\ \, F_L(\omega,q,\theta_d) \,.
\end{equation}
We do not consider other cross section parts, where the transverse current is involved.
The reason is that we intend to compare our results to the data of 
 ref.~\cite{NIKHEF}. In \cite{LOTT1} it was shown that for the kinematics of that experiment
the transverse current contributions to the cross section are small.   
In eq.~(\ref{cross}) $e_{k^\prime}$ and $\Omega_{e^\prime}$ denote energy and solid 
angle of the scattered electron, and $\sigma_M$ is the Mott cross section. Energy 
and momentum transfer to the nuclear system are denoted by 
$\omega$ and ${\bf q}=q\hat{\bf q}$, $q_\mu^2$ is the squared four--momentum 
transfer, $\theta_d$ denotes the angle between ${\bf p}_d=p_d \hat{\bf p}_d$, 
the outgoing deuteron 
momentum, and ${\bf q}$, while $M_d$ is the deuteron mass. Note that for given
values of $\omega$, $q$, and $p_d$ there is a unique deuteron knock--out angle $\theta_d$.
The quantity $F_L(\omega,q,\theta_d)$ is the longitidinal structure function,
defined as
\begin{equation}\label{FL}
{F_L(q, \omega,\theta_d)=  (G_E^p(q_\mu^2))^2 \sum _{M,M'}
\left|\left\langle\Psi_{MM'}^-(E_{d,d})
\left|\hat \rho({\bf q})\right|\Psi_{\alpha}\right\rangle\right|^2 }\,.
\end{equation}
Here $\Psi^-_{MM'}$ is the internal continuum final state of the 
minus type pertaining to the d-d channel with a given relative momentum  
of the final d-d pair (denoted below by ${\bf k}$) and deuteron spin projections $M$ and $M'$.
The quantity $E_{d,d}$ denotes the excitation
energy of the four--nucleon system, 
\begin{equation}\label{Edd} 
E_{d,d} = \epsilon_{d,d}- 2 E_d+E_\alpha   \,\,\, {\rm with} \,\,\, \epsilon_{d,d} = 
{\frac {k^2} {M_d}} \,,
\end{equation} 
where $E_d$ is the deuteron 
binding energy and $E_\alpha$ is the $^4$He  binding energy.
The four--body ground state is 
denoted by $\Psi_{\alpha}$, and
$G_E^p$ is the proton electric form factor.
As nuclear charge operator $\hat \rho$ we take 
\begin{equation}\label{rho}
\hat \rho({\bf q}) =\sum_{j=1}^{4}{
\frac{1+\tau_{j}^{3}}{2}}
\exp{(i {\bf q}\cdot {\bf r}_j}) \,.
\end{equation}
Here $\tau^3_j$ denotes the third component 
of the $j$-th nucleon isospin and ${\bf r}_j$ represents the position of
the $j$-th nucleon with respect to the center of mass of the four--body system.

In our calculation we do not make use of the 
continuum wave function $\Psi_{MM'}^-$, but  instead determine $F_L(q, \omega,\theta_d)$
by means of the integral transform 
method for exclusive reactions \cite{Efros} with the Lorentz kernel 
\cite{LIT,Lapiana}. In this approach one avoids to treat a continuum state problem, 
one works instead with a much easier bound--state--like problem. Our calculation is 
carried out in complete analogy to the $^4$He$(\gamma,p)^3$H and $^4$He$(e,e^\prime p)^3$H
calculations of \cite{Sofia1,Sofia2}, thus in the following we give only a very
brief summary of the method. 

The starting point of the calculation are the transition matrix elements
\begin{equation}\label{Tpt}
T_{MM'}(E_{d,d})=
\left\langle\Psi_{MM'}^-(E_{d,d})\left|\hat \rho\right|
\Psi_{\alpha}\right\rangle~.
\end{equation}
They can be divided into a Born term,
\begin{equation}\label{TBorn}
T_{MM'}^{Born}(E_{d,d})=\left\langle\phi_{MM'}^-(E_{d,d})\left|
{\widehat{\mathcal A}}~{\hat \rho}\right|\Psi_\alpha\right\rangle~,
\end{equation}
and a  FSI dependent term, 
\begin{equation}\label{TFSI1}
T_{MM'}^{FSI}(E_{d,d})=\left\langle\phi_{MM'}^-(E_{d,d})\left|
{\mathcal V}{\widehat{\mathcal A}}\frac{1}{E_{d,d}+i\varepsilon-H} 
\,{\hat \rho}\right|\Psi_\alpha\right\rangle~.
\end{equation}
In the above two equations $\phi_{MM'}^-$ is a product of the internal
wave functions of the two deuterons, with spin projections $M$ and $M'$,
and of the Coulomb function for their relative motion with a 
given large--distance momentum. The relative motion occurs in the 
average d-d Coulomb potential. 
The potential ${\mathcal V}$ is the sum of all interactions 
between nucleons belonging to different deuterons with the average d-d Coulomb potential
being subtracted, $H$ denotes the full 
Hamiltonian of the four--nucleon system, and ${\widehat{\mathcal A}}$
is an antisymmetrization operator explicitly given in ref.~\cite{Diego}.

The calculation of the Born term is rather unproblematic, while the FSI term is 
calculated with the integral transform method. To this end first the following 
identity is used
\begin{eqnarray}\label{TFSI3}                                        
T_{MM'}^{FSI}(E_{d,d})&=&\int_{E_{th}^-}^{\infty}{\frac{F_{MM'}(E)}  
{E_{d,d}+i\varepsilon-E}                                             
dE}~=~-i\pi F_{MM'}(E_{d,d})+{\mathcal P}\int_{E_{th}^-}^{\infty}    
{\frac{F_{MM'}(E)}{E_{d,d}-E}dE}~ \,,                                  
\end{eqnarray} 
where $E_{th}$ is the break--up threshold energy of $^4$He in the isospin $T=0$ channel
and
\begin{equation}\label{F}
F_{MM'}(E)=\sum\!\!\!\!~\!\!\!\!\!\!\int d\nu\left\langle\phi_{MM'}^-
(E_{d,d})\left|{\mathcal V}{\widehat{\mathcal A}}\right|\Psi_\nu(E_\nu)
\right\rangle\left\langle\Psi_\nu(E_\nu)\left|{\hat \rho} \right|\Psi_\alpha
\right\rangle\delta(E-E_\nu)~.
\end{equation}
The function $F_{MM'}$ contains information on all the 
eigenstates $\Psi_\nu$ for the whole eigenvalue spectrum of $H$. In the LIT
method it is obtained by its Lorentz integral transform
\begin{equation}
L\left[F_{MM'}\right](\sigma)=
\int_{E_{th}^-}^{\infty}
{\frac{F_{MM'}(E)}
{(E-\sigma_R)^2+\sigma_I^2}~dE}~=~
\left\langle {\widetilde \Psi}_2(\sigma)
\left|\right.{\widetilde
\Psi}_1(\sigma)\right\rangle~,
\label{mod}
\end{equation}
where
\begin{equation}\label{Psitilda12}
{\widetilde\Psi}_1(\sigma)=(H-\sigma)^{-1}{\hat \rho}\left|\Psi_\alpha
\right\rangle,\qquad
{\widetilde\Psi}_2(\sigma)=(H-\sigma)^{-1}{\widehat {\mathcal A}}
{\mathcal V}
|\phi_{MM^\prime}^-(E_{d,d})\rangle
\end{equation}
and $\sigma=\sigma_R+i\sigma_I$.
Equation (\ref{mod}) shows that $L\left[F_{MM'}\right](\sigma)$ can 
be calculated  without explicit knowledge of $F_{MM'}$, provided that 
one solves the two equations 
\begin{eqnarray}
(H-\sigma)\left|{\widetilde\Psi}_1\right\rangle&=&{\hat \rho}
\left|\Psi_\alpha\right\rangle\label{psitilde1}~,\\
(H-\sigma)\left|{\widetilde\Psi}_2\right\rangle&=&{\widehat 
{\mathcal A}}{\mathcal V}|\phi_{MM'}^-(E_{d,d})\rangle \,.
\label{psitilde2} 
\end{eqnarray}
The quantities ${\widetilde\Psi}_1$ and ${\widetilde\Psi}_2$ have 
finite norms and thus only bound state techniques are required to obtain the
 solutions of eqs.~(\ref{psitilde1}) and (\ref{psitilde2}). 

We use expansions over a basis set of localized functions consisting of correlated 
hyperspherical harmonics (CHH) multiplied by hyperradial functions,
which lead to rather large Hamiltonian matrices. Instead of using a time consuming 
inversion method we directly evaluate the scalar products in (\ref{mod}) with the 
Lanczos technique as explained in ref.~\cite{MBLO}.

After having calculated $L[F_{MM^\prime}](\sigma)$ one obtains the function 
$F_{MM^\prime}(E)$, and thus $T_{MM^\prime}(E_{d,d})$, via the inversion of the LIT. 
We perform the inversion as described in~\cite{ELO99} (for other
inversion methods see \cite{ALRS}).

\section{RESULTS}
\label{Sec:Results}

In our calculation we use the semi--realistic MT-I/III potential \cite{MT} as NN 
interaction. Below pion threshold it leads to rather good descriptions of the NN 
s--wave phase shifts $^3S_1$ and $^1S_0$. The Coulomb interaction is considered in 
addition.  As already mentioned, the ground state of $^4$He as well as
$\tilde \Psi$ of eqs.~(\ref{psitilde1}) and (\ref{psitilde2}) are calculated
using the CHH expansion method.  In order to speed up the convergence, state 
independent correlations are introduced as in~\cite{ELO97} (our $^4$He wave function 
is identical with the CHH $\Psi_\alpha$ of~\cite{BELO}). The deuteron ground state
wave function is determined by a numerical solution of the radial Schr\"odinger
equation. As proton electric form factor we take the usual dipole parametrization.
The transition matrix elements~(\ref{TBorn}) and~(\ref{TFSI1}) are calculated in the
form of partial wave expansions. For the Born term we take into account multipoles
up to order $L=20$, while for the FSI term we include multipoles up to $L=5$.
We checked that with such an expansion a sufficent convergence is reached.

In this work we  consider the $^4$He$(e,e'd)$ reactions measured
in the above mentioned NIKHEF experiment \cite{NIKHEF}. This corresponds
to the following kinematical settings: relative energy of the two 
final deuterons $\epsilon_{d,d}= 35$ MeV
%, missing momentum $p_m=125$ MeV/c, 
and four--momentum transfers $q_\mu^2 = 1.75$, 2.49, 3.36 and 4.79 fm$^{-2}$. 

We start our discussion considering the effect of the Coulomb FSI. In fig.~1
results are shown for the highest of the four considered momentum transfers. As one may expect
the effect is very small and only visible in the cross section minimum around 
$\theta_d = 50$ degrees. Similarly small effects of the Coulomb FSI are
found for the other three momentum transfers. Also shown in fig.~1 is the PWIA 
result of \cite{LOTT2}, where a harmonic oscillator (HO) s--state wave function is
taken for the $^4$He ground state. Although such a ground state model is rather 
different from ours, one finds rather similar results in both calculations for the 
cross section minimum, while at forward and backward angles the HO ground state leads
to an increase of the cross section by about a factor of two.

In fig.~2 we illustrate the effect of the full FSI for all four considered 
momentum transfers. Though the positions of the minima at about 50 degrees
are hardly changed, it is readily evident that FSI is very important
for all four considered momentum transfers. For the lower two q--values one obtains 
reductions of strength by about a factor of 5, while the shapes of the 
angular distribution remain almost unchanged. On the contrary for the two higher 
momentum transfers also the shape is affected. One finds stronger decreases of 
$F_L$ only in the forward region and for deuteron angles beyond 100 degrees 
and accordingly much less pronounced minima. 

In fig.~3 we show the differential cross section resulting from the calculated $F_L$ 
at $q_\mu^2 = 4.79$ fm$^{-2}$. Different from the PWIA case the calculation
with inclusion of FSI does not exhibit a minimum, but shows a rather constant 
fall--off with increasing deuteron knock--out angle. As already seen in fig.~2
the cross section
is significantly reduced by FSI for forward directions and beyond 100 degrees.
In comparison to the experimental cross section at $\theta_d \simeq 15$ degrees 
one finds for the PWIA an overestimation by about a factor of 2.6 and
an underestimation by about a factor of 1.7 for the full calculation.

In fig.~4 we make a comparison of our results with experimental data also for 
the other considered momentum transfers. One sees that the rather strong 
reductions of the cross sections due to FSI is by far not sufficient for an agreement 
with experiment at the lower momentum transfers. In fact the disagreement with 
experiment becomes more and more pronounced with decreasing momentum transfer. 
For the lowest q--value the experimental cross section is overestimated by 
almost an order of magnitude. We mention once again that an inclusion
of the here not considered transverse current contributions should not lead
to significant changes. In fact in ref. \cite{LOTT1} it was shown that the one--body
current increases the PWIA result by less than 20 \% for the lower three momentum
transfers and by about 25 \% for $q_\mu^2=4.79$ fm$^{-2}$. Additional two--body
currents are expected to be small because of the isoscalar nature of the
$^4$He$(e,e'd)d$ reaction. In fig.~4 we also show theoretical results from 
\cite{LOTT1}. A considerably smaller discrepancy between theory and experiment 
is found than in our calculation. It should be pointed out that in \cite{LOTT1} 
FSI is not calculated microscopically as in our case, but modelled via a central 
phenomenological potential between the two outgoing deuterons. As already 
mentioned the $^4$He 
ground state of \cite{LOTT1} is a simple harmonic oscillator wave function.

In order to better understand the origin of the differences between the two 
calculations we show in fig.~5 the corresponding PWIA results. One notices 
that a large part of the differences of fig.~4 are already found for the PWIA 
results and thus are caused by the different ground state wavefunctions. In the 
figure we also illustrate the PWIA calculation of Morita taken from \cite{LOTT1}, 
where correlations are introduced in the $^4$He ground state via the ATMS method 
\cite{ATMS}. The ATMS and HO results are rather similar and considerably lower than 
ours below $q_\mu^2 = 2.5$ fm$^{-2}$. The different results show that it is not 
very likely that a fully realistic calculation of the $^4$He ground state could 
close the gap to experiment at lower momentum transfer. Most probably one needs 
a more realistic description of the FSI for the four--nucleon final state. It
should be noted that the FSI effects of the present work and those of 
\cite{LOTT1} are not very different, in fact in both cases one finds similar 
reductions of the corresponding PWIA cross sections. However, we want to
emphasize once again that there are principal differences for the treatment of FSI
in both calculations. On the other hand both FSI calculations have in common that 
only central potentials between nucleons or respectively deuterons are taken 
into account. Probably a consideration of tensor and other realistic force terms could 
change the picture significantly.

We summarize our work as follows. We have calculated the $^4$He$(e,e'd)d$ 
reaction taking into account the full four--nucleon dynamics in inital and
final states. Such two--body knock--out reactions are considered as an excellent
tool to determine two--body ground state correlations. In fact the comparison
of our PWIA results with corresponding results from other calculations show
a non negligible sensitivity of the cross section to ground state correlations.
On the other hand we find that FSI effects are even more important leading to a
strong  reduction of the cross section. We thus confirm similar results from ref.
\cite{LOTT1}, where, however, FSI was considered in a d-d cluster model only 
and not as in our case microscopically via a NN interaction. Compared to experimental data
our theoretical results show deviations up to about 50 \% in the range
$3.3$ fm$^{-2}$ $\le q_\mu^2  \le 4.8$ fm$^{-2}$, but exhibit  
a considerable overestimation of the experimental cross section at lower momentum     
transfers. As already pointed out a more realistic $^4$He ground state will 
most likely not be sufficient for a satisfying improvement, whereas a more realistic
nuclear interaction also for the final four--nucleon state might lead to 
an agreement of theoretical and experimental results.

%%%%%%%%%%%%%%%%%%%%%%%%%%%%%%%%%%%%%%%%%%%%%%%%%%%%%%%%%%%%%%%%%%%%%%%%%%
%%                             BIBLIOGRAFIA                             %%
%%%%%%%%%%%%%%%%%%%%%%%%%%%%%%%%%%%%%%%%%%%%%%%%%%%%%%%%%%%%%%%%%%%%%%%%%%

\newpage

\begin{figure}
\resizebox*{17cm}{12cm}{\includegraphics{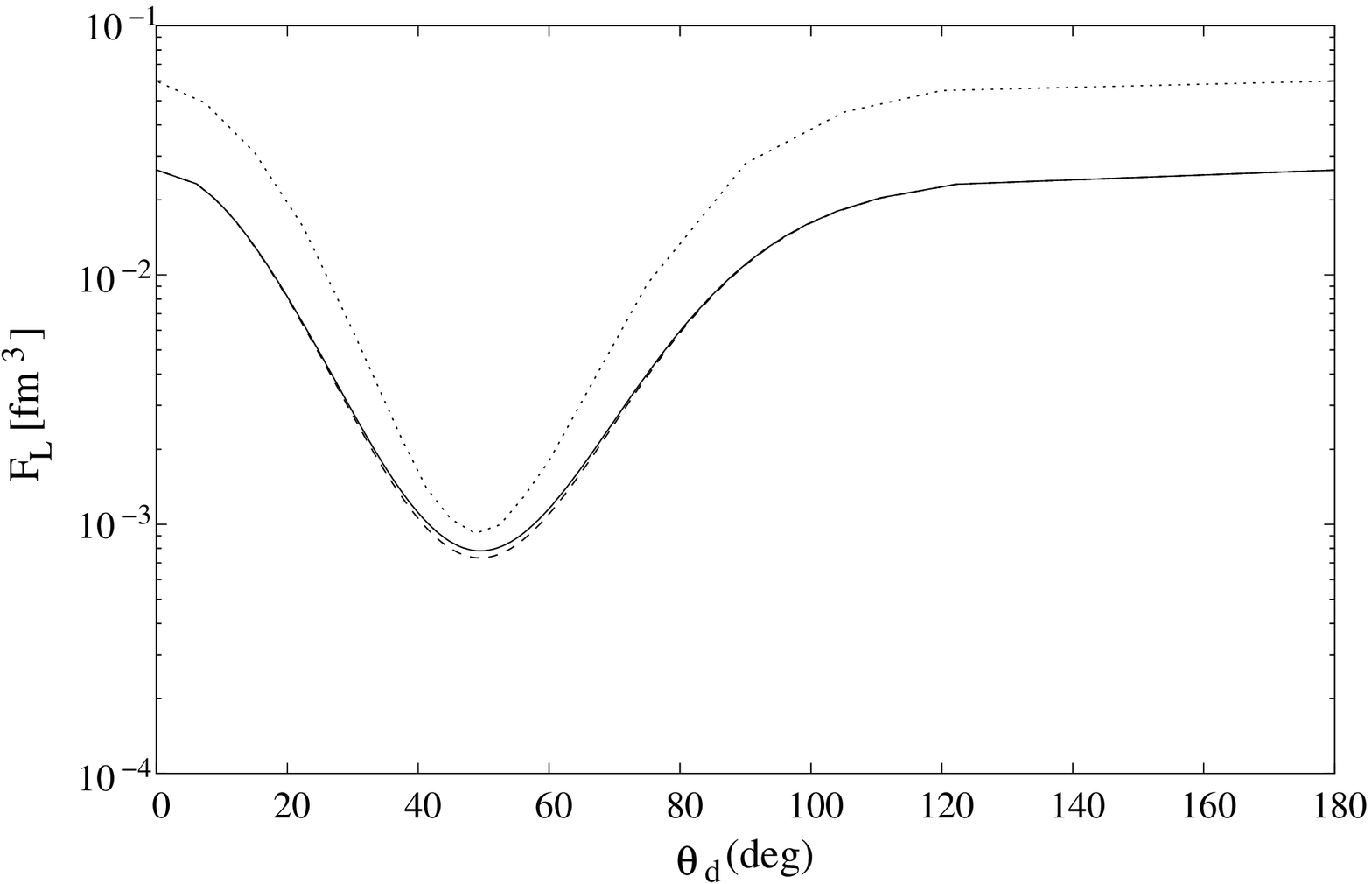}}
\caption{Angular distribution of $F_L$  at $\epsilon_{d,d}=35$ MeV and 
$q_\mu^2=4.79$ fm$^{-2}$ without any FSI (dashed curve) and with
inclusion of Coulomb-FSI (full curve); also shown the PWIA result of \cite{LOTT2}
(dotted curve).}        
\end{figure}      

\begin{figure}
\resizebox*{17cm}{12cm}{\includegraphics{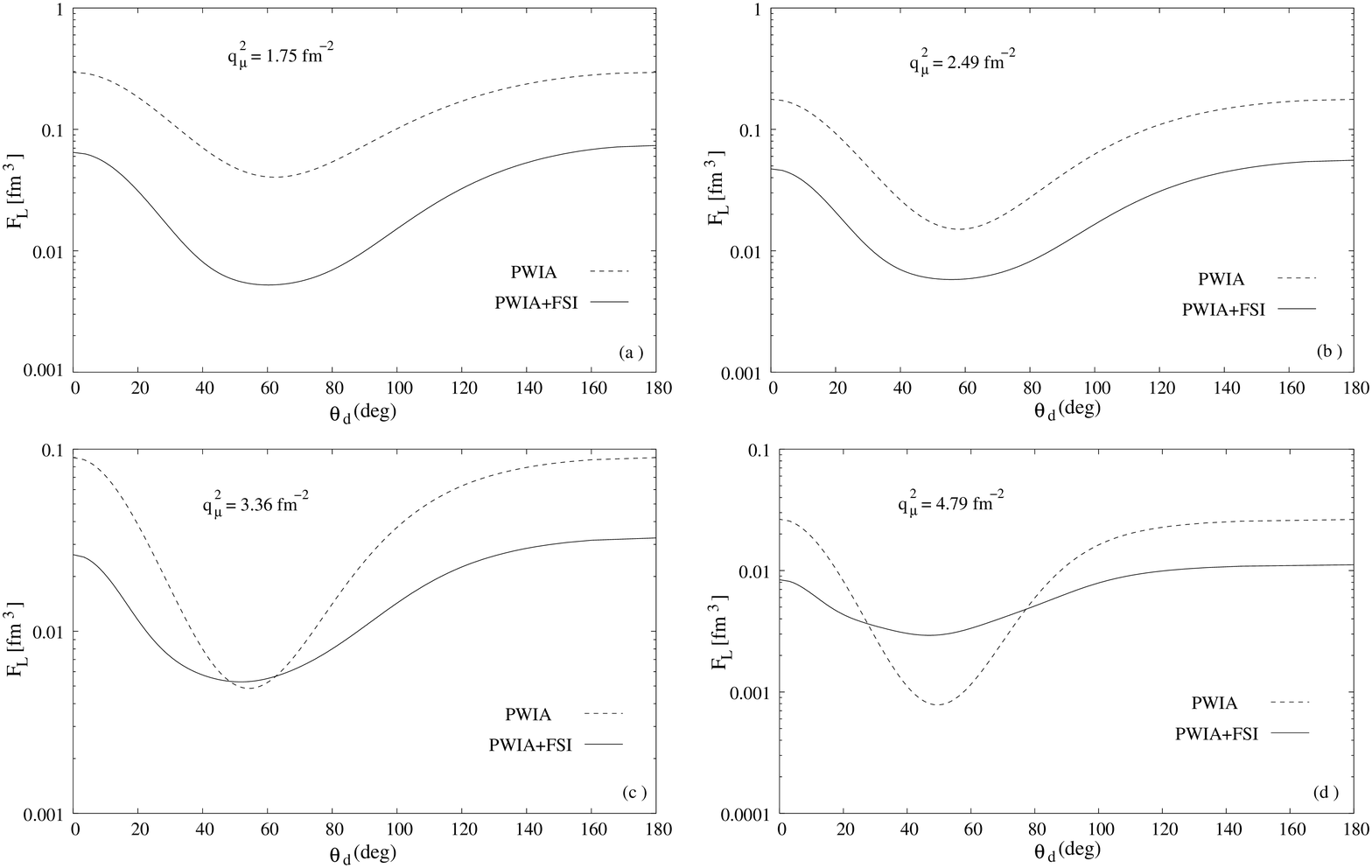}}
\caption{Angular distribution of $F_L$ with (full curve) and without 
(dashed curve) FSI contribution at $\epsilon_{d,d}=35$ MeV and momentum 
transfers as indicated in the figure.}
\end{figure}
                                                                   
\begin{figure}
\resizebox*{17cm}{12cm}{\includegraphics{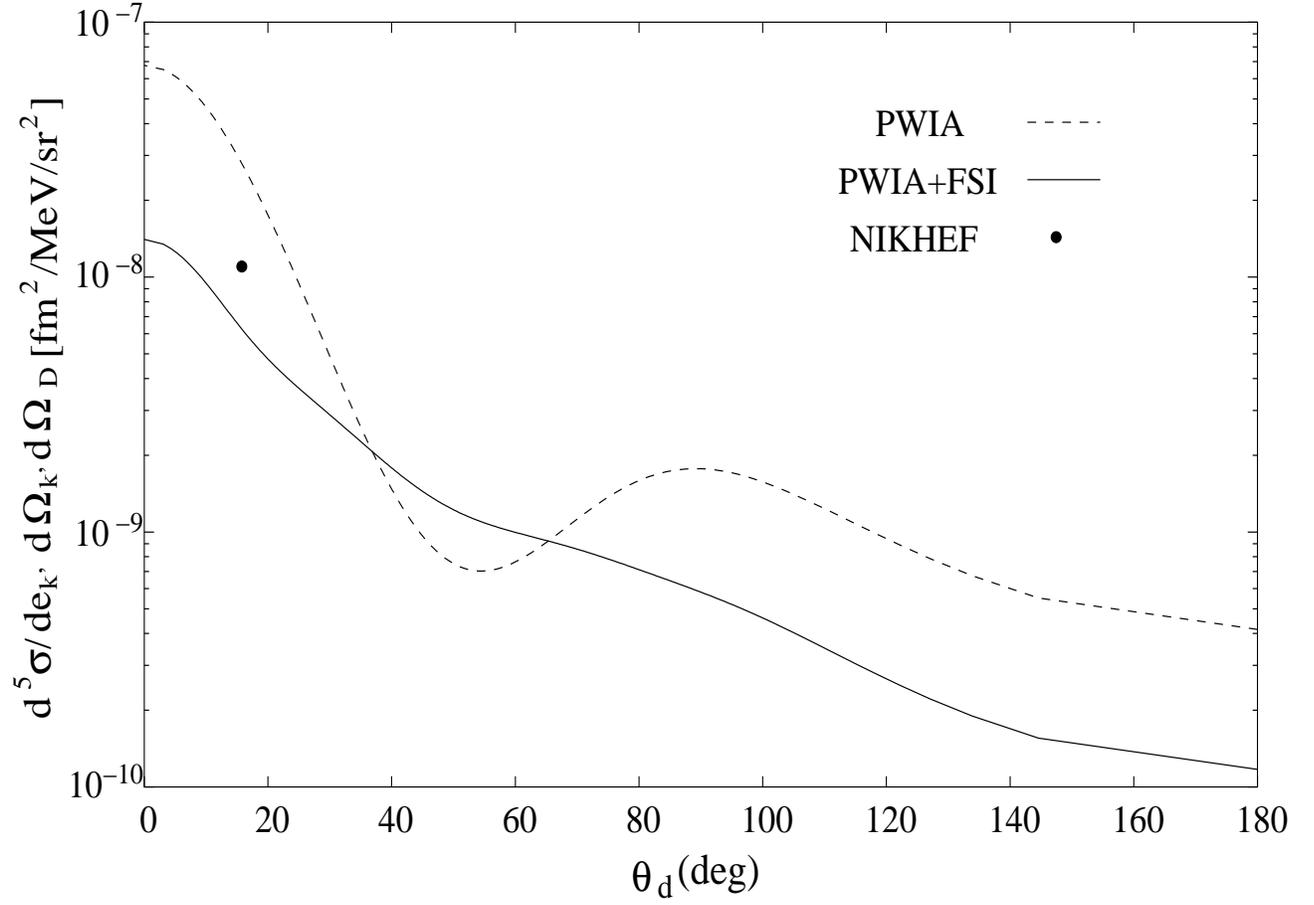}}
\caption{Angular distribution of differential cross section at 
$\epsilon_{d,d}=35$ MeV and $q_\mu^2=4.79$ fm$^{-2}$ with (full curve) 
and without (dashed curve) FSI; also shown experimental result (dot) 
from \cite{NIKHEF}.}
\end{figure}

\begin{figure}
\resizebox*{17cm}{12cm}{\includegraphics{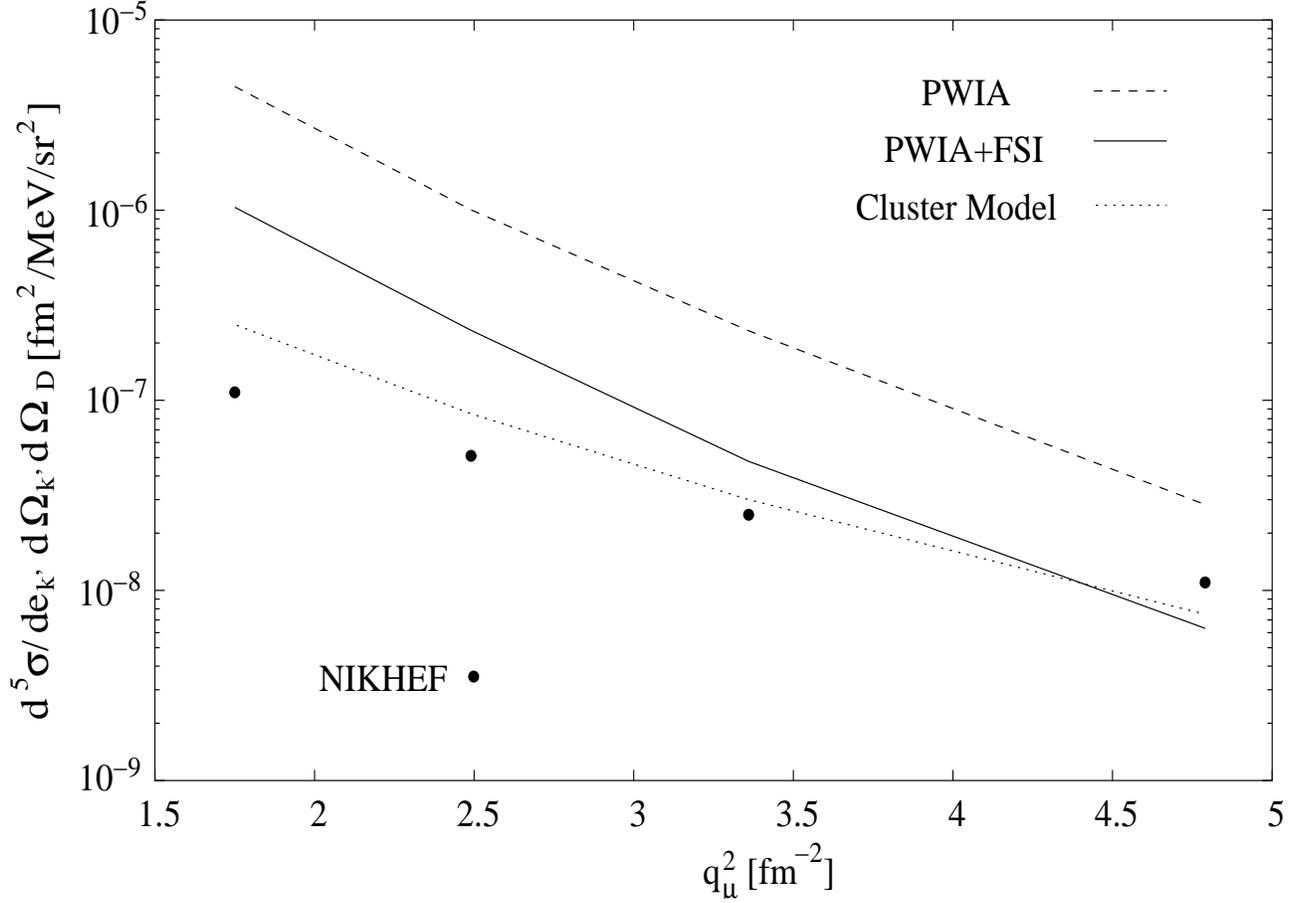}}
\caption{Differential cross section at $\epsilon_{d,d}=35$ MeV and missing 
deuteron momentum ${\bf p}_m={\bf q}-{\bf p}_d$ of $|{\bf p}_m|=125$ MeV/c as 
function of $q_\mu^2$ with (full 
curve) and without FSI (dashed curve) in comparison to experimental data 
(dots) \cite{NIKHEF}; also shown result from \cite{LOTT1} with an HO $^4$He 
ground state and FSI in a d-d cluster model (dotted curve).}       
\end{figure}                                                                   

\begin{figure}
\resizebox*{17cm}{12cm}{\includegraphics{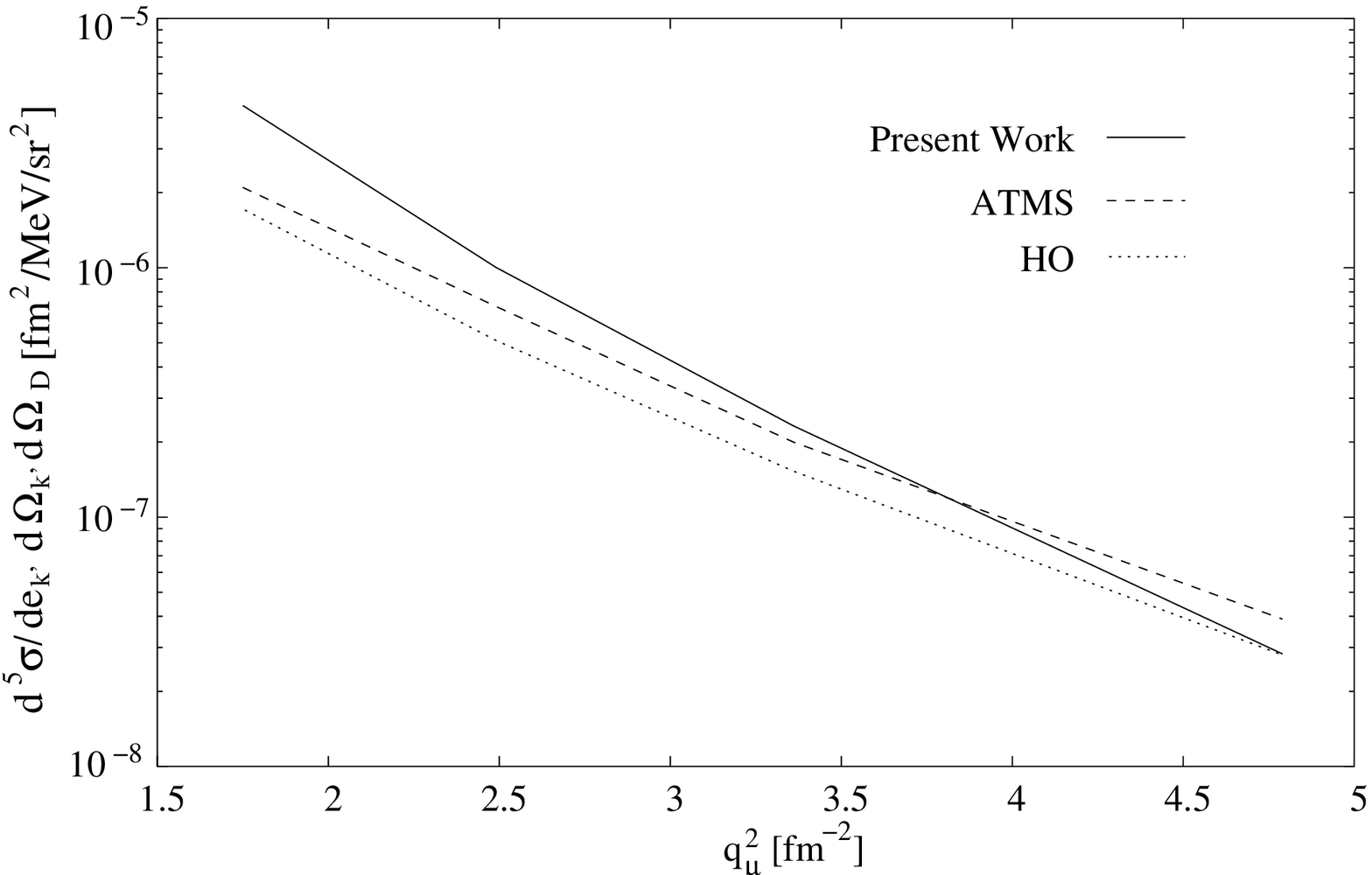}}
\caption{Various PWIA results for the differential cross section 
(kinematics as in fig.~4): present work (full curve), HO \cite{LOTT1}
(dotted curve), and Morita's ATMS from \cite{LOTT1} (dashed curve).} 
\end{figure}    


\begin{thebibliography}{00}
\bibitem{overview} {\it Proceedings of the 6th Workshop on "Electromagnetically
 Induced Two--Hadron Emissions", Pavia 2003}, edited by A. Braghieri, C. Giusti,
P. Grabmayr, ISBN 88-85159-20-6.
\bibitem{ALT} H. Arenh\"ovel, W. Leidemann, E.L. Tomusiak, Eur. Phys. J. A.
 {\bf 23}, 147 (2005).
\bibitem{BoKr} J. Golak, R. Skibinski, H. Witala, W. Gl\"ockle, A. Nogga,
 H. Kamada, Phys. Rep. {\bf 415}, 89 (2005).
\bibitem{LOTT2} W. Leidemann, G. Orlandini, M. Traini, E. L. Tomusiak, 
Phys. Rev. C {\bf 50}, 630 (1994).
\bibitem{NIKHEF} R. Ent, H. Blok, J.F.J. van den Brand, H.J. Bulten, E. Jans,    
L. Lapikas, H. Morita, Phys. Rev. Lett. {\bf67}, 18 (1991).  
\bibitem{LOTT1} W. Leidemann, G. Orlandini, M. Traini, E. Tomusiak, 
Phys. Lett. B {\bf 279}, 212 (1992).
\bibitem{Efros} V.D. Efros, Yad. Fiz. {\bf 41}, 1498 (1985) [Sov. J. Nucl. 
 Phys. {\bf 41}, 949 (1085)], V.D. Efros, Yad. Fiz. {\bf 62}, 1975 (1999) 
 [Phys. Atom. Nucl. {\bf 62}, 1833 (1999)].
\bibitem{LIT}  V. Efros, W. Leidemann, G. Orlandini, Phys. Lett. B
 {\bf 338}, 130 (1994).  
\bibitem{Lapiana}  A. La Piana, W. Leidemann, Nucl. Phys. A {\bf 677}, 423 (2000). 
\bibitem{Sofia1}  S. Quaglioni, N. Barnea, V.D. Efros, W. Leidemann, 
G. Orlandini, Phys. Rev. C {\bf 69}, 044002 (2004).
\bibitem{Sofia2}  S. Quaglioni, V.D. Efros, W. Leidemann, G. Orlandini, 
nucl-th/050842.
\bibitem{Diego} D. Andreasi, Thesis, Universit\`a di Trento (2005).
\bibitem{MBLO} M.A. Marchisio, N. Barnea, W. Leidemann, G. Orlandini,
Few-Body Syst. {\bf 33}, 259 (2003).
\bibitem{ELO99} V.D. Efros, W. Leidemann, G. Orlandini, Few-Body Syst.
{\bf 26}, 251 (1999).
\bibitem{ALRS} D. Andreasi, W. Leidemann, C. Reiss, M. Schwamb, Eur. Phys. J.A.
{\bf 24}, 361 (2005).
\bibitem{MT} R.A. Malfliet, J. Tjon, Nucl. Phys. A {\bf 127}, 161 (1969).
\bibitem{ELO97} V.D. Efros, W. Leidemann, G. Orlandini, Phys. Rev. Lett. 78,
 432 (1997).
\bibitem{BELO} N. Barnea, V.D. Efros, W. Leidemann, G. Orlandini, Phys. Rev. C      
{\bf 63}, 057002 (2001). 
\bibitem{ATMS} H. Morita, Y. Akaishi, H. Tanaka, Prog. Theor. Phys. {\bf 79},
863 (1988).

\end{thebibliography}
\end{document}